%% file: arxiv_version_revised.tex
\documentclass[10pt,twocolumn,letterpaper]{article}

\usepackage{wacv}
\usepackage{times}
\usepackage{epsfig}
\usepackage{graphicx}
\usepackage{amsmath}
\usepackage{amssymb}
\usepackage{subcaption}
\usepackage{enumitem}
\usepackage{authblk}

\def\wacvPaperID{386} 

\wacvfinalcopy 
\ifwacvfinal
\def\assignedStartPage{9876} 
\fi

\ifwacvfinal
\usepackage[breaklinks=true,bookmarks=false]{hyperref}
\else
\usepackage[pagebackref=true,breaklinks=true,colorlinks,bookmarks=false]{hyperref}
\fi

\ifwacvfinal
\else
\fi

\begin{document}

\title{Self-Supervised Poisson-Gaussian Denoising}

\author[1]{Wesley Khademi}
\author[2]{Sonia Rao}
\author[3]{Clare Minnerath}
\author[4]{Guy Hagen}
\author[1]{Jonathan Ventura}
\affil[1]{California Polytechnic State University}
\affil[2]{University of Georgia}
\affil[3]{Providence College}
\affil[4]{University of Colorado Colorado Springs}
\affil[]{\tt\small \{wkhademi, jventu09\}@calpoly.edu}





\maketitle

\begin{abstract}
We extend the blindspot model for self-supervised denoising to handle Poisson-Gaussian noise and introduce an improved training scheme that avoids hyperparameters and adapts the denoiser to the test data.  Self-supervised models for denoising learn to denoise from only noisy data and do not require corresponding clean images, which are difficult or impossible to acquire in some application areas of interest such as low-light microscopy.  We introduce a new training strategy to handle Poisson-Gaussian noise which is the standard noise model for microscope images.  Our new strategy eliminates hyperparameters from the loss function, which is important in a self-supervised regime where no ground truth data is available to guide hyperparameter tuning.  We show how our denoiser can be adapted to the test data to improve performance. Our evaluations on microscope image denoising benchmarks validate our approach.
\end{abstract}

\input{fig-mice-revised}

Fluorescence microscopy is a vital tool for understanding cellular processes and structures. Because fluorescence imaging with long exposure times or intense illumination may damage the cell sample through phototoxicity, fluorescence microscopy images are typically acquired under photon-limited conditions. However, safely imaging the cell using low light conditions and/or low exposure times unfortunately lowers the signal-to-noise ratio (SNR), hindering further analysis and interpretation of the resulting images.

The SNR is the product of a combination of factors, including exposure time, excitation intensity, and camera characteristics.  In fluorescence microscopy, the noise is typically described by a Poisson-Gaussian model \cite{foi2008practical}. The goal of image denoising is to computationally increase the image SNR (Figure \ref{fig:mice}).  In contrast to traditional methods \cite{chambolle2004algorithm,dabov2006image,blu2007sure,luisier2010image,buades2011non,gu2014weighted} which denoise based on only the input image, learning-based methods learn to denoise from a dataset of example images.

In recent years, deep learning methods using convolutional neural networks have shown significant promise in learning-based fluorescence microscopy image denoising \cite{zhang2017beyond,weigert2018content}.  However, the supervised approach to learning denoising faces practical limitations because it requires a large number of corresponding pairs of low SNR and high SNR images.  When imaging live cells, for example, it is not possible to acquire paired low and high SNR images for training because a) the sample is moving and b) exposure to light causes photobleaching and ultimately kills the sample.

For these reasons, researchers have turned to self-supervised approaches to denoising \cite{soltanayev2018training,batson2019noise2self,krull2019noise2void,laine2019high}.  In the self-supervised setting, the learner only has access to low SNR images.  Of the recent approaches, blindspot neural networks \cite{laine2019high} have shown the best performance.  In this work, we address two shortcomings of blindspot neural networks for self-supervised denoising:
\begin{enumerate}
\item We introduce a loss function appropriate for Poisson-Gaussian noise which is the standard model for microscope images;
\item We introduce an alternate training strategy which eliminates the need to regularize the loss function; this is critical in the self-supervised setting where no ground truth validation data is available to tune the regularization strength.
\end{enumerate}

In the following, we survey related work on self-supervised denoising (Section \ref{sec:related_work}), review the blindspot neural network approach to self-supervised denoising (Section \ref{sec:self_supervised}), introduce our new uncalibrated approach (Section \ref{sec:uncalibrated}), present the results of our evaluation and comparison to competing methods on benchmark datasets (Section \ref{sec:experiments}), and provide conclusions and directions for future work (Section \ref{sec:conclusions}).

\section{Related Work}
\label{sec:related_work}

\subsection{Traditional methods}

Many traditional methods for denoising such as BM3D \cite{dabov2006image}, non-local means \cite{buades2011non}, and weighted non-nuclear norm minimizaiton \cite{gu2014weighted} perform denoising by comparing the neighborhood of a pixel to other similar regions in the image.  The advantage of learning-based methods is that they can also take advantage of examples from other images in the dataset beyond the input image to be denoised. Other methods such as total-variation denoising \cite{chambolle2004algorithm} enforce smoothness priors on the image which tend to lead to highly quantized results. 

While most previous methods for denoising are designed for additive Gaussian noise; in the case of Poisson-Gaussian noise, a variance stabilizing transform \cite{makitalo2012optimal} is applied to approximately transform the noise to be Gaussian.  However, these methods are designed explicitly for Poisson-Gaussian noise \cite{luisier2010image}.

\subsection{Deep learning methods}

At present, supervised deep learning methods for denoising \cite{zhang2017beyond,weigert2018content} typically far outperform traditional and self-supervised methods in terms of peak signal-to-noise ratio (PSNR).  Most supervised methods apply a fully convolutional neural network \cite{long2015fully,ronneberger2015u} and simply regress to the clean image.

Recently, several approaches to self-supervised denoising have been developed. 
Some methods \cite{NIPS2018_7587} use as a loss function Stein's Unbiased Risk Estimate (SURE) \cite{stein1981estimation,ramani2008monte}, which estimates the mean squared error (MSE) between a denoised image and the clean image without actually having access to the clean image.  An analogous estimator for Poisson-Gaussian noise has been developed \cite{le2014unbiased}.  However, these methods require \emph{a priori} knowledge of the noise level which is unrealistic in a practical setting.  Our approach supports blind denoising and adaptively estimates the noise level at test time.

Lehtinen et al.~\cite{lehtinen2018noise2noise} introduced a highly successful approach to self-supervised denoising called Noise2Noise.  In this approach, the network learns to transform one noisy instantiation of a clean image into another; under the MSE loss function, the network learns to output the expected value of the data which corresponds to the clean image.  While this method can achieve results very close to a supervised method, it requires multiple, corresponding noisy images and thus is similarly limited in application in the live cell microscopy context.

An alternate approach to self-supervised denoising which does not require multiple noise instantiations of the same clean image is to learn a filter which predicts the center pixel of the receptive field based on the surrounding neighborhood of noisy pixels.  By training such a filter to minimize the MSE to the noisy input, the resulting filter will theoretically output the clean value \cite{batson2019noise2self,krull2019noise2void}.  Laine et al.~\cite{laine2019high} refer to a neural network built around this concept as a ``blindspot neural network.''  They improved upon the blindspot concept by extending it to a Bayesian context and introduced loss functions for pure Gaussian or Poisson noise, showing results very close to the supervised result when trained on synthetically noised data.  However, their method requires a regularization term in the loss function which can't practically be tuned in the self-supervised setting; in our evaluation we found that the regularization strength indeed needs to be tuned for best results on different datasets.  Our method avoids the need for regularization and outperforms the regularized version in our experiments.

Krull et al.~\cite{krull2019probabilistic} introduced Probabilistic Noise2Void (PN2V) which takes a non-parametric approach to modeling both the noise distribution and the network output; however, their approach requires paired clean and noisy images in order to calibrate the noise model.  A recent follow-on work called PPN2V \cite{prakash2019fully} estimates the noise model using a Gaussian Mixture Model (GMM) in a fully unsupervised manner.  Again, this approach involves several hyperparameters controlling the complexity of the noise model which need to be tuned, while ours does not.  Additionally, in our experiments, we show that our approach outperforms PPN2V on several datasets.

\section{Self-supervised learning of denoising}
\label{sec:self_supervised}

The goal of denoising is to predict the values of a ``clean'' image $\mathbf{x} = (x_1,\ldots,x_n)$ given a ``noisy'' image $\mathbf{y}=(y_1,\ldots,y_n)$.  
Let $\Omega_{y_i}$ denote the neighborhood of pixel $y_i$, which does not include $y_i$ itself.  We make two assumptions that are critical to the setup of Noise2Void \cite{krull2019noise2void} and follow-on works: that the noise at each pixel is sampled independently, i.e. $p(y_i|x_1,\ldots,x_n)=p(y_i|x_i)$; and that each clean pixel is dependent on its neighborhood, a common assumption about natural images.  The consequence of these assumptions is that $\Omega_{y_i}$ only gives information about $x_i$, not $y_i$.  Therefore a network trained to predict $y_i$ given $\Omega_{y_i}$ using a mean squared error loss will in fact learn to predict $x_i$ \cite{krull2019noise2void,batson2019noise2self}. 

In this work, we take a probabilistic approach rather than trying to regress to a single value.  Following Laine et al.~\cite{laine2019high} and Krull et al.~\cite{krull2019probabilistic}, we can connect $y_i$ to its neighborhood $\Omega_{y_i}$ by marginalizing out the unknown clean value $x_i$:
\begin{equation}
    \underbrace{p(y_i|\Omega_{y_i})}_{\text{Noisy observation}} = \int \underbrace{p(y_i|x_i)}_{\text{Noise model}}\underbrace{p(x_i|\Omega_{y_i})}_{\text{Clean prior}} dx_i.
\end{equation}
Since we only have access to observations of $y_i$ for training, this formulation allows us to fit a model for the clean data by minimizing the negative log likelihood of the noisy data, i.e. minimizing a loss function defined as
\begin{equation}
\label{eq:loss}
\mathcal{L}^{\text{marginal}} = \sum_i -\log p(y_i|\Omega_{y_i}).
\end{equation}
In the following we will drop the $\Omega_{y_i}$ to save space.

\subsection{Poisson-Gaussian noise}

In the case of Poisson-Gaussian noise, the noisy observation $y_i$ is sampled by first applying Poisson corruption to $x_i$ and then adding Gaussian noise which is independent of $x_i$. We have
\begin{equation}
\label{eq:poisson_gaussian_noise}
y_i = aP(x_i/a) + N(0,b)
\end{equation}
where $a>0$ is a scaling factor (related to the gain of the camera) and $b$ is the variance of the Gaussian noise component, which models other sources of noise such as electric and thermal noise \cite{foi2008practical}.

We apply the common approximation of the Poisson distribution as a Gaussian with equal mean and variance:
\begin{align}
\label{eq:pg_noise}
y_i &\approx aN(x_i/a,x_i/a) + N(0,b) \\
&= N(x_i,a x_i + b).
\end{align}
The noise model is then simply a Gaussian noise model whose variance is an affine transformation of the clean value.  Note that in practice we allow $b$ to be negative; this models the effect of an offset or ``pedestal'' value in the imaging system \cite{foi2008practical}.  This general formulation encompasses both pure Gaussian ($a=0$) and Poisson noise ($b=0$).

\subsection{Choice of prior}

In order to implement our loss function (Equation \ref{eq:loss}) we need to choose a form for the prior $p(x_i|\Omega_{y_i})$ that makes the integral tractable.  One approach is to use the conjugate prior of the noise model $p(y_i|x_i)$, so that the integral can be computed analytically.  For example, Laine et al.~\cite{laine2019high} model the prior $p(x_i|\Omega_{y_i})$ as a Gaussian, so that the marginal is also a Gaussian.  Alternatively, Krull et al.~\cite{krull2019probabilistic} take a non-parametric approach and sample the prior.

In this work, similar to Laine et al.~\cite{laine2019high} we model the prior as a Gaussian with mean $\mu_i$ and variance $\sigma^2_i$.  We replace the $ax$ term in Equation \ref{eq:pg_noise} with $a\mu$ to make the integral in Equation \ref{eq:loss} tractable; this approximation should be accurate as long as $\sigma^2_i$ is small.  The marginal distribution of $y_i$ is then
\begin{equation}
p(y_i)=
\frac{1}{\sqrt{2\pi(a\mu_i+b+\sigma_i^2)}}\exp \left( -\frac{(y_i-\mu_i)^2}{2(a\mu_i+b+\sigma_i^2)}\right)
\end{equation}
and the corresponding loss function is
\begin{equation}
\mathcal{L}^{\text{marginal}} = \sum_i \left( \frac{(y_i-\mu_i)^2}{a\mu_i+b+\sigma_i^2}+\log(a\mu_i+b+\sigma_i^2) \right)
\end{equation}

\subsection{Posterior mean estimate}

At test time, $\mu_i$ is an estimate of the clean value $x_i$ based on $\Omega_{y_i}$, the neighborhood of noisy pixels around $y_i$.  However, this estimate does not take into account the actual value of $y_i$ which potentially provides useful information about $x_i$.

Laine et al.~\cite{laine2019high} and Krull et al.~\cite{krull2019probabilistic} suggest to instead use the expected value of the posterior to maximize the PSNR of the resulting denoised image.  In our case we have
\begin{equation}
\label{eq:pme}
\hat{x}_i =
\mathbb{E}[p(x_i|y_i)] =
\frac{y_i \sigma_i^2 + (a\mu_i+b) \mu_i}{a\mu_i + b + \sigma_i^2}.
\end{equation}
Intuitively, when the prior uncertainty is large relative to the noise estimate, the formula approaches the noisy value $y_i$; when the prior uncertainty is small relative to the noise estimate, the formula approaches the prior mean $\mu_i$.

\subsection{Blindspot neural network}
In our approach, $\mu_i$ and $\sigma_i^2$ are the outputs of a blindspot neural network \cite{laine2019high} and $a$ and $b$ are global parameters learned along with the network parameters.

The ``blind-spot neural network'' is constructed in such a way that the network cannot see input $y_i$ when outputting the parameters for $p(x_i)$.  The blindspot effect can be achieved in multiple ways.  Noise2Void ~\cite{krull2019noise2void} and Noise2Self \cite{batson2019noise2self} replace a random subset of pixels in each batch and mask out those pixels in the loss computation.  Laine et al.~\cite{laine2019high} instead construct a fully convolutional neural network in such a way that the center of the receptive field is hidden from the neural network input.  In our experiments we use the same blindspot neural network architecture as Laine et al.~\cite{laine2019high}.

\subsection{Regularization}

\input{table-loss}

\input{table-synthetic-params}

In a practical setting, the parameters $a$ and $b$ of the noise model are not known \emph{a priori}; instead, we need to estimate them from the data.  However, an important issue arises when attempting to learn the noise parameters along with the network parameters: the network's prior uncertainty and noise estimate are essentially interchangeable without any effect on the loss function.  In other words, the optimizer is free to increase $a$ and $b$ and decrease $\sigma_i^2$, or vice-versa, without any penalty.  To combat this, we add a regularization term to the per-pixel loss which encourages the prior uncertainty to be small:
\begin{equation}
\mathcal{L}^{\text{regularized}} = \mathcal{L}^{\text{marginal}} + \lambda \sum_i |\sigma_i|.
\end{equation}

We found in our experiments that the choice of $\lambda$ strongly affects the results.   When $\lambda$ is too high, the prior uncertainty is too small, and the results are blurry.  When $\lambda$ is too low, the prior uncertainty is too high, and the network does not denoise at all.  Unfortunately, in the self-supervised setting, it is not possible to determine the appropriate setting of $\lambda$ using a validation set, because we do not have ground truth ``clean'' images with which to evaluate a particular setting of $\lambda$.

\section{Learning an uncalibrated model}
\label{sec:uncalibrated}

This realization led us to adopt a different training strategy which defers the learning of the noise parameter models to test time.  

In our uncalibrated model, we do not separate out the parameters of the noise model from the parameters of the prior.  Instead, we learn a single variance value $\hat{\sigma_i}^2$ representing the total uncertainty of the network.  Our uncalibrated loss function is then
\begin{equation}
\label{eq:uncalib_loss}
\mathcal{L}^{\text{uncalibrated}}=\sum_i
\left(
\frac{(y_i-\mu_i)^2}{\hat{\sigma_i}^2}+\log(\hat{\sigma_i}^2)
\right)
\end{equation}

At test time, however, we need to know the noise parameters $a$ and $b$ in order to compute $\sigma_i^2=\hat{\sigma_i}^2-a\mu_i-b$ and ultimately compute our posterior mean estimate $\hat{x}_i$.  

If we had access to corresponding clean and noisy observations ($x_i$ and $y_i$, respectively) then we could fit a Poisson-Gaussian noise model to the data in order to learn $a$ and $b$.  In other words, we would find
\begin{equation}
\label{eq:pg_fit}
a,b = {\arg \min}_{a,b} \sum_i \left( \frac{(y_i-x_i)^2}{ax_i+b}+\log(ax_i+b)\right).
\end{equation}

As we are in a self-supervised setting, however, we do not have access to clean data.  Instead, we propose to use the prior mean $\mu_i$ as a stand-in for the actual clean value $x_i$.  This bootstrapping approach is similar to that proposed by Prakash et al.~\cite{prakash2019fully}; however, they fit a general parametric noise model to the training data where as we propose to fit a Poisson-Gaussian model to each image in the test set.

Our approach is summarized in the following steps:
\begin{enumerate}
    \item Train a blindspot neural network to model the noisy data by outputting a mean and variance value at each pixel, using the uncalibrated loss (Equation \ref{eq:uncalib_loss}).
    \item For each test image:
    \begin{enumerate}[label=\roman*.]
        \item Run the blindspot neural network with the noisy image as input to obtain mean $\mu_i$ and total variance $\hat{\sigma_i}^2$ estimate at each pixel.
        \item Determine the optimal noise parameters $a,b$ by fitting a Poisson-Gaussian distribution to the noisy and psuedo-clean images given by the mean values of the network output (Equation \ref{eq:pg_fit}).
        \item Calculate the prior uncertainty at each pixel as $\sigma_i^2 = \max(0.0001,\hat{\sigma_i}^2-a\mu_i-b)$.
        \item Use the noise parameters $a,b$ and the calculated prior uncertainties $\sigma_i^2$ to compute the denoised image as the posterior mean estimate (Equation \ref{eq:pme}).
    \end{enumerate}
\end{enumerate}

We believe our approach has two theoretical advantages over the bootstrap method proposed by Prakash et al. \cite{prakash2019fully}.  We can achieve a better fit to the data by training our system end-to-end, whereas Prakash et al. \cite{prakash2019fully} impose a fixed noise model during training by first estimating the noise parameters and then training the network. Second, we estimate noise parameters for each image separately at test time, whereas Prakash et al. \cite{prakash2019fully} estimate common noise parameters for all images first and fixes those parameters during training.  Our approach allows for slight deviations in the noise parameters for each image, which might be more realistic for an actual microscope imaging system where the camera configuration slightly fluctuates between images.

\input{fig-hist-fmd}

\section{Experiments and Results}
\label{sec:experiments}

\subsection{Implementation details}

Our implementation uses the Keras library with Tensorflow backend.  We use the same blindspot neural network architecture as Laine et al.~\cite{laine2019high}.  We use the Adam optimizer \cite{kingma2014adam} with a learning rate of 0.0003 over 300 epochs, halving the learning rate when the validation loss plateaued.  Each epoch consists of 50 batches of $128\times128$ crops from random images from the training set.  For data augmentation we apply random rotation (in multiples of 90 degrees) and horizontal/vertical flipping.

To fit the Poisson-Gaussian noise parameters at test time, we apply Nelder-Mead optimization \cite{nelder1965simplex} with $(a=0.01,b=0)$ as the initialization point.  We cut off data in the bottom 2\% and top 3\% of the noisy image's dynamic range before estimating the noise parameters.

\subsection{Datasets}

\subsubsection{Synthetic Data}

We generate a synthetic dataset using the ground truth images of the Confocal MICE dataset from the FMD benchmark \cite{zhang2019poisson} (described below). For training, we use the ground truth images from 19 of the 20 views and generate 50 noisy examples of each view by synthetically adding Poisson-Gaussian noise to the ground truth images using equation \ref{eq:poisson_gaussian_noise} where $a=1/\lambda$ and $b=(\sigma/255)^2$. For testing, we use the ground truth image from the 20th view and generate 50 noisy examples by synthetically adding Poisson-Gaussian noise in the same manner as during training. To ensure our method works for a wide range of noise levels, we train/test our method on all combinations $(\lambda,\sigma) \in \{0,10,20,30,40,50\}\times\{0,10,20,30,40,50\}$. 

\input{table-synthetic-psnr}

\subsubsection{Real Data}

We evaluated our method on two datasets consisting of real microscope images captured with various imaging setups and types of samples.  Testing on real data gives us a more accurate evaluation of our method's performance in contrast to training and testing on synthetically noised data, since real data is not guaranteed to follow the theoretical noise model.  

The fluoresence microscopy denoising (FMD) benchmark \cite{zhang2019poisson} consists of a total of 12 datasets of images captured using either a confocal, two-photon, or widefield microscope.  We used the same subset of datasets (Confocal Mice, Confocal Fish, and Two-Photon Mice) used to evaluate PN2V \cite{krull2019probabilistic} so that we could compare our results.  Each dataset consists of 20 views of the sample with 50 noisy images per view.  The 19th-view is withheld for testing, and the ground truth images are created by averaging the noisy images in each view.  We trained a denoising model on the raw noisy images in each dataset separately.

Prakash et al.~\cite{prakash2019fully} evaluated PPN2V on three sequences from a confocal microscope, imaging Convallaria, Mouse Nuclei, and Mouse Actin.  Each sequence consists of 100 noisy images and again the clean image is computed as the average of the noisy images.  Whereas the FMD dataset provides 8-bit images clipped at 255, these images are 16-bit and thus are not clipped.  Following their evaluation procedure, each method is trained on all 100 images and then tested on a crop of the same 100 images; this methodology is allowable in the self-supervised context since no label data is used during training. 

\input{table-reg}

\subsection{Experiments}

In the following we will refer to the competing methods under consideration as
\begin{itemize}
    \item \textbf{Regularized (Ours)}: Blindspot neural network trained using the regularized Poisson-Gaussian loss function (Equation \ref{eq:loss}) with regularization strength $\lambda$.  
    \item \textbf{Uncalibrated (Ours)}: Blindspot neural network trained using the uncalibrated loss function (Equation \ref{eq:uncalib_loss}) with noise parameter estimation done adaptively at test time (Section \ref{sec:uncalibrated}).
    \item \textbf{N2V}: Noise2Void which uses the MSE loss function and random masking to create the blindspot effect \cite{krull2019noise2void}.
    \item \textbf{PN2V}: Probabilistic Noise2Void -- same setup as N2V but uses a histogram noise model created from the ground truth data and a non-parametric prior \cite{krull2019probabilistic}.
    \item \textbf{Bootstrap GMM} and \textbf{Bootstrap Histogram}: PPN2V training -- same setup as PN2V but models the noise distribution using either a GMM or histogram fit to the Noise2Void output \cite{prakash2019fully}.
    \item \textbf{U-Net}: U-Net \cite{ronneberger2015u} trained for denoising in a supervised manner using MSE loss \cite{weigert2018content}.
    \item \textbf{N2N}: Noise2Noise training using MSE loss \cite{lehtinen2018noise2noise}.
\end{itemize}

\input{table-results}
\input{fig-images-fmd-revised}
\input{fig-images}

\subsubsection{Noise parameter estimation}

We first evaluate whether our bootstrap approach to estimating the Poisson-Gaussian noise parameters is accurate in comparison to estimating the noise parameters using the actual ground truth clean values.


To evaluate our bootstrapping method, we compare the ground truth and estimated Poisson-Gaussian noise models fit for a test image in each dataset in the FMD benchmark \cite{zhang2019poisson}. Figure \ref{fig:hist-fmd} shows that the Poisson-Gaussian pdfs generated using our bootstrapping technique closely match that of the Poisson-Gaussian pdfs generated from the ground truth images.

We further evaluate our approach by comparing the loss and estimated Poisson-Gaussian noise parameters obtained when using actual ground truth data or the pseudo-clean data generated in our bootstrap method. Table \ref{table:loss} shows that bootstrapping can provide an accurate estimation of noise parameters and result in a loss similar to that obtained from using ground truth clean data.  Here the loss value is
\begin{equation}
\label{eq:pg_fit}
\frac{1}{N} \sum_i \left( \frac{(y_i-x_i)^2}{ax_i+b}+\log(ax_i+b)\right)
\end{equation}
where $y_i$ is a pixel from the noisy image and $x_i$ is a corresponding pixel from either the ground truth clean image or the pseudo-clean image.

We perform a similar evaluation on our synthetic dataset where instead of having to estimate the true noise parameters from fitting a Poisson-Gaussian noise model with the ground truth clean image we readily have available the true noise parameters that correspond to the level of synthetically added Poisson-Gaussian noise. Table \ref{table:synthetic-params} shows the true noise parameters as well as the ones obtained using our uncalibrated method and the method described by Foi et al. in \cite{foi2008practical}. Our method tends to overestimate the $a$ parameter, whereas the estimate of the $b$ parameter is consistently accurate. This is probably because a majority of the pixels in the Confocal MICE images are dark and thus there are not many good samples for fitting the level of Poisson noise, whereas every pixel can be effectively used to estimate the Gaussian noise no matter the underlying brightness. Unlike the method of Foi et al.~\cite{foi2008practical} which obtains poor noise estimates most likely because of this, our method is still able to obtain a good estimate of the parameters by leveraging information from both the noisy image and our pseudo-clean image.

The effectiveness of fitting a Poisson-Gaussian noise model at test time is further evaluated in Table \ref{table:synthetic-psnr} which provides a comparison of peak signal-to-noise ratio (PSNR) on a subset of our synthetic dataset. Our method of estimating the noise parameters with our bootstrapping technique consistently improves the denoised results of the pseudo-clean image, but is ultimately bounded by the result obtained from using the pseudo-clean image along with the true noise parameters.  Results for all noise parameter combinations are given in the supplemental material.

\subsubsection{Effect of regularization}

To highlight the difficulties of hyperparameter tuning in the self-supervised context, we trained our uncalibrated model and several regularized models on the FMD datasets.  We tested a regularization strength of $\lambda=0.1,1,$ and $10$.

The results are shown in Table \ref{table:reg}.  The test set PSNR of the regularized model varies greatly depending on the setting of $\lambda$, and indeed a different setting of $\lambda$ is optimal for each dataset.  This indicates that hyperparameter tuning is critical for the regularized approach, but it is not actually possible in a self-supervised context.  

In contrast, our uncalibrated method outperforms the regularized method at any setting of $\lambda$, and does not require any hyperparameters.

\subsubsection{Comparison to state-of-the-art}
Next we present the results of our performance evaluation on the FMD and PPN2V benchmark datasets.   Table \ref{table:results} shows a comparison between our uncalibrated method and various competing methods, including self-supervised and supervised methods.  

Between the fully unsupervised methods that do not require paired noisy images (our Uncalibrated method, N2V, Bootstrap GMM, and Bootstrap Histogram), our method outperforms the others on four out of six datasets. A comparison of denoising results on both benchmark datasets are shown in Figures \ref{fig:images-fmd} and \ref{fig:images}.

\section{Conclusions and Future Work}
\label{sec:conclusions}

Noise is an unavoidable artifact of imaging systems, and for some applications such as live cell microscopy, denoising is a critical processing step to support quantitative and qualitative analysis.  In this work, we have introduced a powerful new scheme for self-supervised learning of denoising which is appropriate for processing of low-light images.  In contrast to the state-of-the-art, our model handles Poisson-Gaussian noise which is the standard noise model for most imaging systems including digital microscopes.  In addition, we eliminate the need for loss function regularization in our method, thus making self-supervised denoising more practically applicable.  Our evaluation on real datasets show that our method outperforms competing methods in terms of the standard PSNR metric on many datasets tested.

Our work opens up new avenues in live-cell imaging such as extreme low-light imaging over long periods of time.  Future work lies in extending our model to other noise models appropriate to other imaging modalities, and exploring whether our uncalibrated method could be combined with a non-parametric prior \cite{krull2019probabilistic}. 

\paragraph{Acknowledgments} This work was supported in part by NSF \#1659788, NIH \#1R15GM128166-01 and the UCCS Biofrontiers Center.

{\small
\bibliographystyle{ieee_fullname}
\bibliography{egbib}
}

\pagebreak

\appendix
\section{Derivation of Poisson-Gaussian Loss Function}
\setcounter{equation}{0}
\setcounter{figure}{0}
\setcounter{table}{0}

In self-supervised denoising, we only have access to noisy pixels $y_i$ and not the corresponding clean pixels $x_i$.  Similar to a generative model, we use the negative log-likelihood of the training data as our loss function:
\begin{equation}
\mathcal{L}_i = - \log p(y_i)
\end{equation}

However, for denoising we are interested in learning a model for $p(x_i)$, not $p(y_i)$.  We relate $p(x_i)$ to $p(y_i)$ by marginalizing out $x_i$ from the joint distribution:
\begin{align}
p(y_i) &= \int_{-\infty}^{\infty} p(y_i,x_i) dx_i \\
&= \int_{-\infty}^{\infty} p(y_i|x_i) p(x_i) dx_i
\end{align}
In other words, we integrate $p(y_i,x_i)=p(y_i|x_i)p(x_i)$ over all possible values of the clean pixel $x_i$.

Here, $p(y_i|x_i)$ is simply our chosen noise model.  $p(x_i)$ constitutes our prior belief about the value of $x_i$ before we have seen an observation of $y_i$.  We do not know what form the prior should take; it is essentially up to us to choose.  Usually we use the conjugate prior of the noise model because this makes the integral tractable.

Our loss function term for pixel $i$ will then be
\begin{equation}
\mathcal{L}_i = -\log \int_{-\infty}^{\infty} p(y_i|x_i) p(x_i) dx_i
\end{equation}

\subsection{Gaussian noise}

For zero-centered Gaussian noise, $p(y_i|x_i)$ is the normal distribution centered at $x_i$ with variance equal to $\sigma_n^2$.    We choose $p(x_i)$ to be the normal distribution as well.  Here we have the network output the parameters of the Gaussian, mean $\mu_i$ and std. dev. $\sigma_i$.

The marginalized pdf is derived as follows:
\begin{align}
p(y_i) &= \int_{-\infty}^{\infty}p(y_i|x_i)p(x_i) dx_i \\
 {}
\begin{split}
= \int_{-\infty}^{\infty}
\frac{1}{\sqrt{2\pi \sigma_n^2}} \exp \left(-\frac{(y_i-x_i)^2}{2\sigma_n^2} \right)  \cdot
\\
\frac{1}{\sqrt{2\pi \sigma_i^2}}
\exp \left( -\frac{(x_i-\mu_i)^2}{2\sigma_i^2} \right) dx_i 
\end{split}
\\
&=  \frac{1}{\sqrt{2\pi(\sigma_n^2+\sigma_i^2)}}\exp \left( -\frac{(y_i-\mu_i)^2}{2(\sigma_n^2+\sigma_i^2)}\right)
\end{align}
which we recognize as a Gaussian with mean $\mu_i$ and variance $\sigma_i^2+\sigma_n^2$.

The loss function is then
\begin{align}
\mathcal{L}_i &= -\log p(y_i) \\
&= -\log \left( \frac{1}{\sqrt{2\pi(\sigma_n^2+\sigma_i^2)}}\exp \left( -\frac{(y_i-\mu_i)^2}{2(\sigma_n^2+\sigma_i^2)}\right) \right) \\
&= \frac{1}{2}\frac{(y_i-\mu_i)^2}{(\sigma_n^2+\sigma_i^2)} + \frac{1}{2}\log 2\pi + \frac{1}{2}\log(\sigma_n^2+\sigma_i^2).
\end{align}
Dropping the constant terms we have
\begin{equation}
\mathcal{L}_i = \frac{(y_i-\mu_i)^2}{(\sigma_n^2+\sigma_i^2)} + \log(\sigma_n^2+\sigma_i^2).
\end{equation}

\subsection{Poisson noise}

For high enough values of $X_i$, the Poisson distribution $\mathcal{P}(\lambda)$ can be approximated by a Gaussian $\mathcal{N}(\lambda,\lambda)$ with mean and variance equal to $\lambda$.  Using this idea, Laine et al.~\cite{laine2019high} adapt the above formulation for Gaussian noise to the Poisson noise case.  However, they introduce an approximation in order to evaluate the integral.

Let $a$ be the scaling factor s.t. $y/a \sim \mathcal{P}(x/a)$ where $x$ and $y$ are in the range $[0~1]$.   The noise model using a normal approximation is $y = a(x/a+N(0,x/a))=x+N(0,ax)$.  The proper joint distribution for this model is thus
\begin{equation}
\begin{split}
p(y_i)p(x_i) = 
\frac{1}{\sqrt{2\pi (a x_i)}}\exp \left(-\frac{(y_i-x_i)^2}{2(a x_i)} \right) \cdot \\
 \frac{1}{\sqrt{2\pi \sigma_i^2}}\exp \left( -\frac{(x_i-\mu_i)^2}{2\sigma_i^2} \right)
\end{split}
\end{equation}
However, Laine et al.~replace the variance of the noise distribution with $a \mu_i$.  This makes the integral tractable.  They argue that this approximation is okay if $\sigma_i^2$ is small.
\begin{align}
\begin{split}
P(y_i) \approx \int_{-\infty}^{\infty}
\frac{1}{\sqrt{2\pi (a \mu_i)}} \exp \left(-\frac{(y_i-x_i)^2}{2(a \mu_i)} \right) \cdot \\
\frac{1}{\sqrt{2\pi \sigma_i^2}}\exp \left( -\frac{(x_i-\mu_i)^2}{2\sigma_i^2} \right)dx_i 
\end{split}
\\
&= \frac{1}{\sqrt{2 \pi (a\mu_i+\sigma_i^2)}} \cdot 
\exp\left( -\frac{(y_i-\mu_i)^2}{2 (a \mu_i+\sigma_i^2)} \right) 
\end{align}
which we recognize as a Gaussian with mean $\mu_i$ and variance $\sigma_i^2+a\mu_i$.

Following the derivation above, our loss function is
\begin{equation}
\mathcal{L}_i = \frac{(y_i-\mu_i)^2}{(a\mu_i +\sigma_i^2)} + \log(a\mu_i+\sigma_i^2).
\end{equation}

\subsection{Poisson-Gaussian noise}

Noise in microscope images is generally modeled as a Poisson-Gaussian process.  The number of photons entering the sensor during the exposure time is assumed to follow a Poisson distribution, and other noise components such as the readout noise and thermal noise are captured by an additive Gaussian term.

We can easily extend the Poisson loss function above to Poisson-Gaussian by adding a noise variance $b$ to the model.  Following the derivation above, our loss function is
\begin{equation}
\mathcal{L}_i = \frac{(y_i-\mu_i)^2}{(a\mu_i +b+\sigma_i^2)} + \log(a\mu_i+b+\sigma_i^2).
\end{equation}

\section{Posterior mean estimate}
The blind-spot network ignores the actual measured value for $y_i$ when it makes a prediction for $x_i$.  However, $y_i$ contains extra information which can be used to improve our estimate of $x_i$.

Laine et al.~\cite{laine2019high} suggest to use the expected value of the posterior:
\begin{align*}
    \hat{x_i} = \mathbb{E}[x_i|y_i] &= \int_{-\infty}^{\infty} p(x_i | y_i)x_i dx_i\\
     &= \frac{1}{Z}\int_{-\infty}^{\infty} p(y_i | x_i)p(x_i)x_i dx_i
\end{align*}
where we have applyed Baye's rule to relate $p(x_i|y_i)$ and $p(y_i|x_i)p(x_i)$ up to a normalizing constant $Z$, where
\begin{equation}
    Z = \int_{-\infty}^{\infty}p(y_i|x_i)dx_i.
\end{equation}

For a Gaussian with prior mean $\mu_i$ and variance $\sigma_i^2$, and noise variance $\sigma_n^2$, we have the following result:
\begin{equation}
\hat{x_i} = \frac{y_i \sigma_i^2 + \sigma_n^2 \mu_i}{\sigma_i^2+\sigma_n^2}
\end{equation}
This same formula can be used for the Poisson or Poisson-Gaussian noise models (replacing $\sigma_n^2$ with $a \mu_i$ or $a\mu_i+b$, respectively).

\section{Results}
\subsection{Gaussian noise}
While our method is meant for denoising images with Poisson-Gaussian noise, we also test our methods ability to denoise pure Gaussian noise. We evaluate two variants of our noise estimation method by comparing the estimated $a$,$b$ parameters to the ground truth parameters for every noise level, $\sigma \in \{10,20,30,40,50\}$, of our synthetic Gaussian noise Confocal MICE dataset. When we can make the assumption that Gaussian noise is the type of noise that exists in an image, we show that an improvement can be made in denoising quality by making a small change to our method to fit pure Gaussian noise parameters. 

Table \ref{table:gaussian-params} provides results obtained on our synthetic dataset in which \emph{Estimated} represents $a$,$b$ parameters obtained using our Poisson-Gaussian noise fitting, while \emph{G-Estimated} represents our estimated $a$,$b$ parameters using our Gaussian noise fitting method which can be accomplished by simply fixing $a=0$ for our Poisson-Gaussian noise fitting technique. While our Poisson-Gaussian noise fitting technique typically provides a better estimate of the $b$ parameter than the Gaussian noise fitting method, our Poisson-Gaussian noise fitting obtains an incorrect estimate of $a$ unlike our Gaussian noise fitting. This worse estimate of $a$ contributes to an overall worse denoising quality than the quality obtained by estimating the noise parameters with Gaussian noise fitting which is shown in Table \ref{table:gaussian-psnr}. This is because the Gaussian noise fitting parameters result in a more accurate estimate of the computed noise variance due to the $a$ parameter being known.

\subsection{Poisson noise}
We perform the same denoising experiments on images with varying levels of Poisson noise. Once again, we show that we can improve the denoising quality by modifying our Poisson-Gaussian noise fitting technique to fit Poisson noise when we can make the assumption that pure Poisson noise is present in an image. This can be done by simply fixing $b=0$ for our Poisson-Gaussian noise fitting technique.

Table \ref{table:poisson-params} shows a comparison between ground truth and estimated $a$,$b$ noise parameters for every noise level, $\lambda \in \{10,20,30,40,50\}$, of our synthetic Poisson noise Confocal MICE dataset. \emph{Estimated} and \emph{P-Estimated} represent our estimated $a$,$b$ parameters using our Poisson-Gaussian and Poisson noise fitting techniques, respectively. While our Poisson-Gaussian noise fitting provides decent estimates of the $a$,$b$ parameters, using the pure Poisson noise fitting can provide almost exact estimates of both parameters. Table \ref{table:poisson-params} shows that our Poisson noise fitting technique results in a PSNR close to that obtained using the ground truth parameters. While the Poisson-Gaussian noise fitting results in a worse denoising quality than the Poisson noise fitting, an improvement in denoising quality is still obtained over the pseudo-clean image.
 
\subsection{Poisson-Gaussian noise}
To further evaluate our bootstrapping method, we compare our estimated $a$,$b$ parameters to the ground truth ones for every noise level, $(\lambda,\sigma) \in \{10,20,30,40,50\}\times\{10,20,30,40,50\}$, of our synthetic Poisson-Gaussian noise Confocal MICE dataset. Table \ref{table:synthetic-params-full} shows the results obtained for all the Poisson-Gaussian noise levels. The ground truth $a$,$b$ parameters correspond to the $(\lambda,\sigma)$ noise level synthetically added to the clean images while the estimated $a$ and $b$ are the parameters learned from fitting a Poisson-Gaussian noise model using our bootstrapping technique.

We further evaluate our approach on our synthetic dataset by comparing the PSNR obtained by our bootstrapping method for all the noise level combinations. Table \ref{table:synthetic-psnr-full} compares the PSNR achieved by the pseudo-clean image which is the output of the blindspot neural network, the image obtained by our bootstrapping method, as well as the image obtained when the ground truth $a$,$b$ parameters are known rather than learned.

\subsection{Noise Parameter Estimation}
To further study how effective our bootstrapping technique is we evaluate how the PSNR is affected by various levels of error in our estimates of the $a$,$b$ parameters. Since the error in our estimation of the $a$ parameter is different than the error in our estimation of the $b$ parameter we select different percent errors for each parameter. The initial percent errors are chosen by computing the average percent error of our estimation of the noise parameters over all noise levels, $(\lambda,\sigma) \in \{0,10,20,30,40,50\}\times\{0,10,20,30,40,50\}$, of our synthetic Confocal MICE dataset. We then select the other two percent errors to be one standard deviation below and above the average percent errors. 

Table \ref{table:psnr-param-error} provides results of the PSNR obtained for the different percent errors in the $a$,$b$ parameters on a Confocal MICE dataset with synthetically added Poisson-Gaussian noise ($\lambda/\sigma = 30$). As mentioned before, our estimation of the $a$ parameter is much worse than the $b$ parameter most likely due to the Confocal MICE dataset containing many dark pixels. Even when using the worst percent errors for our $a$,$b$ parameters though we still obtain an increase in PSNR from the pseudo-clean image which shows that even a poor estimation using our method can still help denoise an image. We also note that the PSNR actually increases with an increase in percent error in the estimate of the $b$ parameter, which can best be explained by our method always overestimating the $a$ parameter and underestimating the $b$ parameter. In practice, a worse estimate of the $b$ parameter helps compensate for the overestimation of the $a$ parameter, providing a better approximation of the noise variance. 

\subsection{BSD68}
We also provide results for our method on the BSD68 dataset. We follow the experiment in \cite{krull2019noise2void} and train our model on 400 gray scale images and test on the gray scale version of the BSD68 dataset. Both the training and testing dataset have synthetically added zero mean Gaussian noise with standard deviation $\sigma = 25$.

We evaluate our method on the BSD68 dataset by comparing the PSNR obtained using BM3D, a supervised training method, Noise2Noise, Noise2Void, and three different outputs of our model: the blindspot neural network, the output obtained estimating the noise parameters, and the output obtained using the true noise parameters. While our method is not expected to perform better than the supervised training method or Noise2Noise, it should perform better than Noise2Void. Table \ref{table:bsd68} compares the different outputs of our model to the various state-of-the-art methods. We observe that the PSNR of the output of the blindspot neural network, or pseudo-clean image, is significantly worse than the PSNR of the output of the Noise2Void method, but when our Poisson-Gaussian noise fitting technique is used we outperform Noise2Void. Similar to Noise2Void, our method performs worse than BM3D which is most likely due to using too small of a training set to obtain the best possible pseudo-clean image from the blindspot neural network.

We further evaluate our method by comparing our estimated $a$,$b$ parameters to the ground truth noise parameters. Table \ref{table:bsd68-params} shows that our method obtains only a decent estimate of the $a$,$b$ parameters. As mentioned, this is probably due to the pseudo-clean image not being close enough to the ground truth clean image to provide a good estimate of the ground truth noise parameters from our Poisson-Gaussian noise fitting technique.

\input{table-gaussian-params}
\input{table-gaussian-psnr}
\input{table-poisson-params}
\input{table-poisson-psnr}

\input{table-synthetic-params-full}
\input{table-synthetic-psnr-full}
\input{table-psnr-param-error}

\input{table-bsd68}
\input{table-bsd68-params}


\end{document}

%% file: fig-mice-revised.tex
\begin{figure}[t]
\label{fig:twophotonmice}
\centering
\begin{subfigure}{.20\textwidth}
    \centering
    \includegraphics[width=0.90\textwidth]{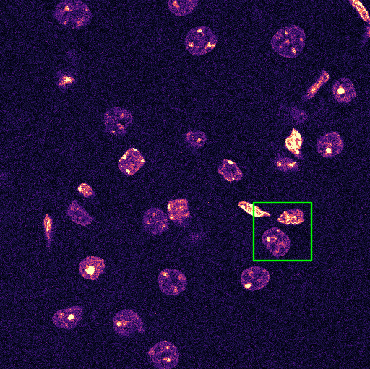}
    \caption{Noisy Input}
\end{subfigure}
\begin{subfigure}{.20\textwidth}
    \centering
    \includegraphics[width=0.90\textwidth]{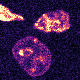}
    \caption{Zoomed Input}
\end{subfigure} \\ [1ex]
\begin{subfigure}{.20\textwidth}
    \centering
    \includegraphics[width=0.90\textwidth]{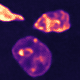}
    \caption{Denoised}
\end{subfigure}
\begin{subfigure}{.20\textwidth}
    \centering
    \includegraphics[width=0.90\textwidth]{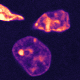}
    \caption{Ground Truth}
\end{subfigure}
\caption[short]{\label{fig:mice}An example of our self-supervised denoising result.  Image from the Confocal Mice dataset \cite{zhang2019poisson}.}
\end{figure}

%% file: table-loss.tex
\begin{table*}
\centering
\begin{tabular}{l |c c| c c |c c} 
 \hline
 & Ground Truth & Bootstrap & Ground Truth & Bootstrap & Ground Truth & Bootstrap \\
 Datasets & \multicolumn{2}{c|}{Loss} & \multicolumn{2}{c|}{$a$}  & \multicolumn{2}{c}{$b$} \\ [0.5ex] 
 \hline\hline
 Confocal Mice & -6.322 & -6.257 & 0.0181 & 0.0196 & -0.000203 & -0.000232 \\ 
 Confocal Fish & -5.224 & -5.122 & 0.0753 & 0.0723 & -0.00101 & -0.00084 \\ 
 Two-Photon Mice & -5.005 & -4.979 & 0.0301 & 0.0296 & -0.000559 & -0.000491 \\ 
[1ex]
 \hline
\end{tabular}
\caption[short]{\label{table:loss}Quantitative comparison of fitting a Poisson-Gaussian noise model using the ground truth clean data or the denoised estimate from the prior. Values in the table are averages over the 50 test images from each dataset.}
\end{table*}

%% file: table-synthetic-params.tex
\begin{table*}
\centering
\begin{tabular}{c | c | c c c | c c c} 
 \hline
 & & Ground Truth & Estimated & Foi et al. \cite{foi2008practical} & Ground Truth & Estimated & Foi et al. \cite{foi2008practical} \\
 $\lambda$ & $\sigma$ & \multicolumn{3}{c|}{$a$}  & \multicolumn{3}{c}{$b$} \\ [0.5ex] 
 \hline\hline
 50 & 10 & 0.0200 & 0.0250 & 0.0252 & 0.00153 & 0.00135 & 0.000900 \\ 
 40 & 20 & 0.0250 & 0.0325 & 0.0759 & 0.00615 & 0.00586 & 0.000334 \\ 
 30 & 30 & 0.0333 & 0.0430 & 0.0971 & 0.0138 & 0.0134 & 0.00108 \\ 
 20 & 40 & 0.0500 & 0.0623 & 0.213 & 0.0246 & 0.0241 & 0.000373 \\ 
 10 & 50 & 0.100 & 0.117 & 0.313 & 0.0384 & 0.0378 & 0.00462 \\ 
[1ex]
 \hline
\end{tabular}
\caption[short]{\label{table:synthetic-params}Quantitative comparison of fitting a Poisson-Gaussian noise model on different noise levels of our synthetic Confocal MICE dataset. \emph{Ground truth} $a$,$b$ parameters correspond to the Poisson-Gaussian noise levels added to our synthetic dataset, \emph{Estimated} represents the $a$,$b$ parameters obtained using our bootstrapping technique, and \emph{Foi et al.} represents the $a$,$b$ parameters obtained using the noise estimation method in \cite{foi2008practical}.}
\end{table*}

%% file: fig-hist-fmd.tex
\begin{figure*}
\centering
\begin{subfigure}{.30\textwidth}
    \centering
    \caption*{Confocal Mice}
    \includegraphics[width=0.95\textwidth]{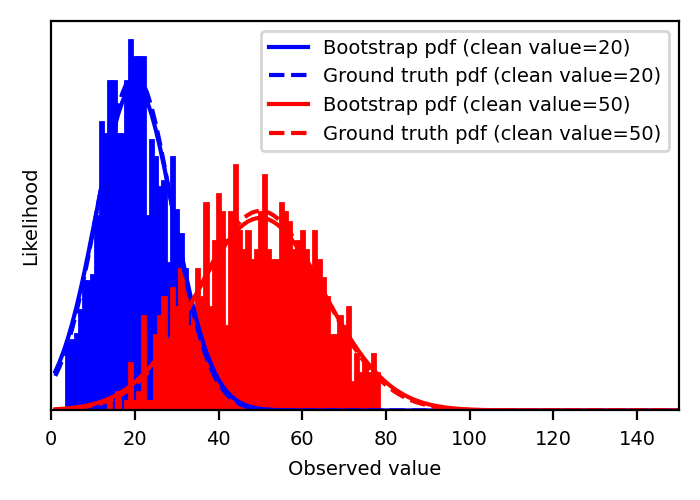}
\end{subfigure}
\begin{subfigure}{.30\textwidth}
    \centering
    \caption*{Confocal Fish}
    \includegraphics[width=0.95\textwidth]{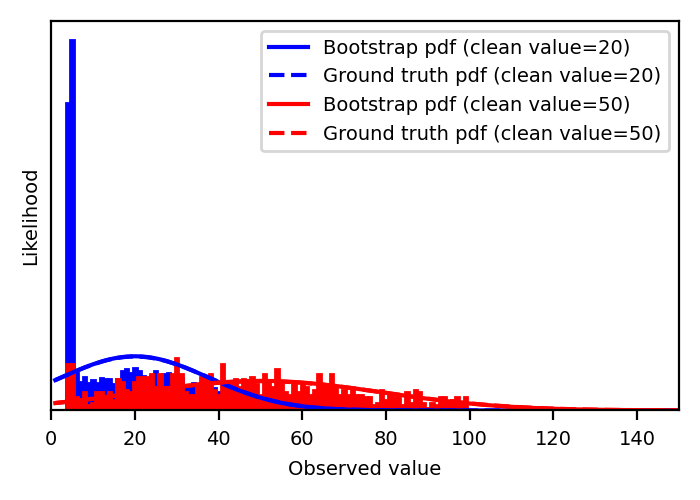}
\end{subfigure}
\begin{subfigure}{.30\textwidth}
    \centering
    \caption*{Two-Photon Mice}
    \includegraphics[width=0.95\textwidth]{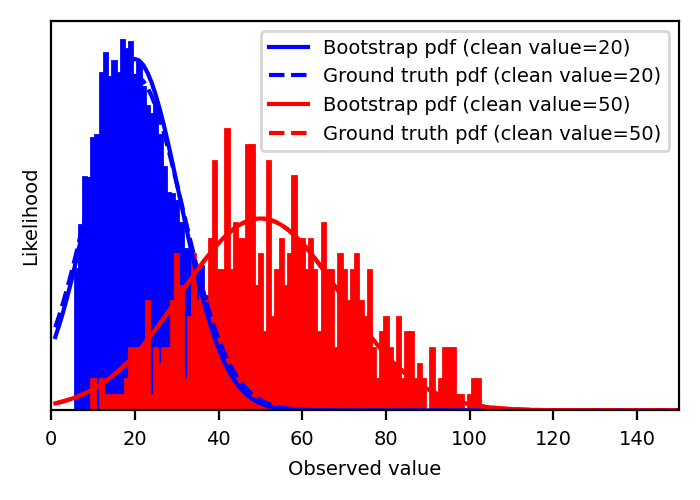}
\end{subfigure}
\caption[short]{\label{fig:hist-fmd}Comparison of Poisson-Gaussian noise models fit to a noisy image from several datasets.  Solid bars show histograms of the noisy values corresponding to a clean value of 20 (blue) and 50 (red).  Curves show the pdfs of a Poisson-Gaussian distribution fit to the data using either ground truth clean data (dashed line) or pseudo-clean data (solid line).  }
\end{figure*}

%% file: table-synthetic-psnr.tex
\begin{table}
\centering
\begin{tabular}{c | c | c | c | c } 
 \hline
 & & Pseudo-Clean & Uncalibrated & Ground Truth \\
 $\lambda$ & $\sigma$ & PSNR & PSNR  & PSNR \\ [0.5ex] 
 \hline\hline
 50 & 10 & 36.95 & 37.05 & 37.25 \\ 
 40 & 20 & 35.45 & 35.53 & 35.73 \\ 
 30 & 30 & 34.10 & 34.15 & 34.33 \\ 
 20 & 40 & 33.17 & 33.22 & 33.37 \\ 
 10 & 50 & 31.98 & 32.03 & 32.14 \\ 
[1ex]
 \hline
\end{tabular}
\caption[short]{\label{table:synthetic-psnr}PSNR comparison on different noise levels of our synthetic Confocal MICE dataset. \emph{Pseudo-clean} is the result obtained before computing the posterior. \emph{Uncalibrated} is the result from computing the posterior with the estimated noise parameters obtained from our bootstrapping technique. \emph{Ground truth} represents computing the posterior with the true $a$,$b$ parameters that correspond to the Poisson-Gaussian noise levels added into our synthetic dataset.}
\end{table} 

%% file: table-reg.tex
\begin{table}[t]
\centering
\begin{tabular}{l c c c c} 
 \hline
 & & Confocal & Confocal & Two-Photon \\
 Methods & $\lambda$ & Mice & Zebrafish & Mice \\ [0.5ex] 
 \hline\hline
 Uncalibrated & - & \textbf{37.97} & \textbf{32.26} & \textbf{33.83} \\
 Regularized & 0.1 & 37.74 & 23.97 & 33.52 \\ 
 Regularized & 1 & 37.64 & 27.44 & 33.56 \\ 
 Regularized & 10 & 37.13 & 31.99 & 33.34 \\ 
[1ex]
 \hline
\end{tabular}
\caption[short]{\label{table:reg}Comparison between uncalibrated and regularized methods.}
\end{table}

%% file: table-results.tex
\begin{table*}[t]
\centering
\begin{tabular}{l c c c c c c c} 
 \hline
 & Confocal & Confocal & Two-Photon & & & \\
 Methods & Mice & Zebrafish & Mice  & Convallaria & Mouse Nuclei & Mouse Actin \\ [0.5ex] 
 \hline\hline
 Uncalibrated (Ours) & \textbf{37.97} & \textbf{32.26} & \textbf{33.83} & 36.44 & \textbf{36.97} & 33.35 \\ 
 N2V & 37.56 & 32.10 & 33.42 & 35.73 & 35.84 & 33.39 \\
 Bootstrap GMM & 37.86 & * & 33.77 & \textbf{36.70} & 36.43 & \textbf{33.74} \\
 Bootstrap Histogram & 36.98 & 32.23 & 33.53 & 36.19 & 36.31 & 33.61 \\ [1ex]
 \hline
 PN2V & 38.24 & 32.45 & 33.67 & 36.51 & 36.29 & 33.78 \\
 U-Net & 38.38 & 32.93 & 34.35 & 36.71 & 36.58 & 34.20 \\ 
 N2N & 38.19 & 32.93 & 34.33 & - & - & - \\ [1ex] 
 \hline
\end{tabular}
\caption{Quantitative comparison of our implementation and baseline methods on datasets provided by Zhang et al.~~\cite{zhang2019poisson} and Prakash et al.~\cite{prakash2019fully}.  Methods above the solid line are fully unsupervised while those below it either require ground truth data or a noisy image pair.  Bold numbers indicate the best performing method among the fully unsupervised methods.  The $*$ indicates a case where the Bootstrap GMM method failed to train (the loss became NaN before convergence).}
\label{table:results}
\end{table*}

%% file: fig-images-fmd-revised.tex
\begin{figure*}
\centering
\raisebox{-0.1in}{\rotatebox[origin=c]{90}{\parbox{2cm}{\centering Confocal Mice}}}
\begin{subfigure}{.13\textwidth}
    \centering
    \caption*{Noisy Input}
    \includegraphics[width=0.98\textwidth]{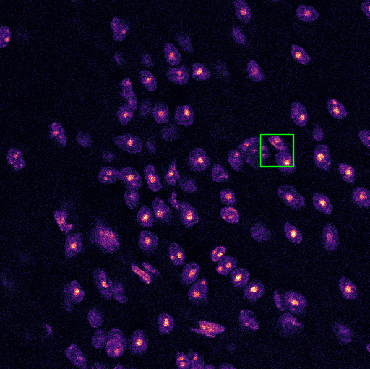}
\end{subfigure}
\begin{subfigure}{.13\textwidth}
    \centering
    \caption*{Zoomed Input}
    \includegraphics[width=0.98\textwidth]{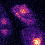}
\end{subfigure}
\begin{subfigure}{.13\textwidth}
    \centering
    \caption*{N2V}
    \includegraphics[width=0.98\textwidth]{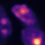}
\end{subfigure}
\begin{subfigure}{.13\textwidth}
    \centering
    \caption*{Bootstrap Hist.}
    \includegraphics[width=0.98\textwidth]{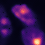}
\end{subfigure}
\begin{subfigure}{.13\textwidth}
    \centering
    \caption*{Bootstrap GMM}
    \includegraphics[width=0.98\textwidth]{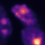}
\end{subfigure}
\begin{subfigure}{.13\textwidth}
    \centering
    \caption*{\textbf{Uncalibrated}}
    \includegraphics[width=0.98\textwidth]{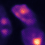}
\end{subfigure}
\begin{subfigure}{.13\textwidth}
    \centering
    \caption*{Ground Truth}
    \includegraphics[width=0.98\textwidth]{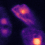}
\end{subfigure} \\
\rotatebox[origin=c]{90}{\parbox{2cm}{\centering Confocal \\Zebrafish}}
\begin{subfigure}{.13\textwidth}
    \centering
    \includegraphics[width=0.98\textwidth]{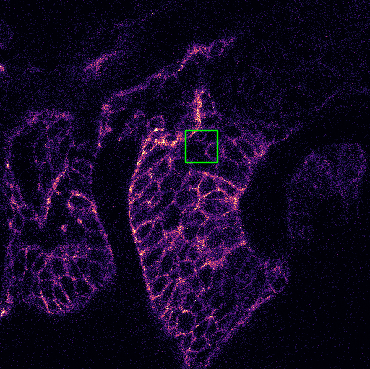}
\end{subfigure}
\begin{subfigure}{.13\textwidth}
    \centering
    \includegraphics[width=0.98\textwidth]{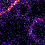}
\end{subfigure}
\begin{subfigure}{.13\textwidth}
    \centering
    \includegraphics[width=0.98\textwidth]{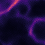}
\end{subfigure}
\begin{subfigure}{.13\textwidth}
    \centering
    \includegraphics[width=0.98\textwidth]{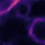}
\end{subfigure}
\begin{subfigure}{.13\textwidth}
    \centering
    \includegraphics[width=0.98\textwidth]{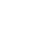}
\end{subfigure}
\begin{subfigure}{.13\textwidth}
    \centering
    \includegraphics[width=0.98\textwidth]{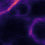}
\end{subfigure}
\begin{subfigure}{.13\textwidth}
    \centering
    \includegraphics[width=0.98\textwidth]{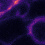}
\end{subfigure} \\
\rotatebox[origin=c]{90}{\parbox{2cm}{\centering Two-Photon Mice}}
\begin{subfigure}{.13\textwidth}
    \centering
    \includegraphics[width=0.98\textwidth]{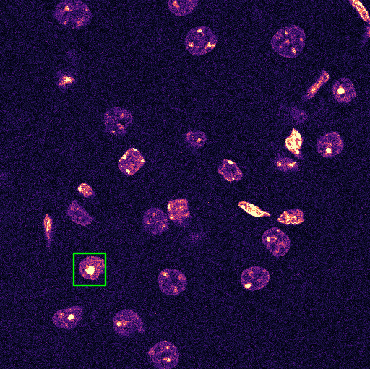}
\end{subfigure}
\begin{subfigure}{.13\textwidth}
    \centering
    \includegraphics[width=0.98\textwidth]{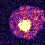}
\end{subfigure}
\begin{subfigure}{.13\textwidth}
    \centering
    \includegraphics[width=0.98\textwidth]{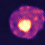}
\end{subfigure}
\begin{subfigure}{.13\textwidth}
    \centering
    \includegraphics[width=0.98\textwidth]{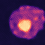}
\end{subfigure}
\begin{subfigure}{.13\textwidth}
    \centering
    \includegraphics[width=0.98\textwidth]{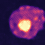}
\end{subfigure}
\begin{subfigure}{.13\textwidth}
    \centering
    \includegraphics[width=0.98\textwidth]{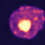}
\end{subfigure}
\begin{subfigure}{.13\textwidth}
    \centering
    \includegraphics[width=0.98\textwidth]{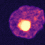}
\end{subfigure}
\caption[short]{\label{fig:images-fmd}Denoising results on images taken from the FMD dataset \cite{zhang2019poisson}. The missing image corresponds to the case where the Bootstrap GMM method failed to train.}
\end{figure*}

%% file: fig-images.tex
\begin{figure*}
\centering
\raisebox{-0.1in}{\rotatebox[origin=c]{90}{Convallaria}}
\begin{subfigure}{.13\textwidth}
    \centering
    \caption*{Noisy Input}
    \includegraphics[width=0.98\textwidth]{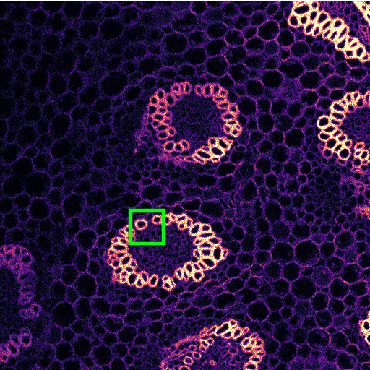}
\end{subfigure}
\begin{subfigure}{.13\textwidth}
    \centering
    \caption*{Zoomed Input}
    \includegraphics[width=0.98\textwidth]{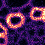}
\end{subfigure}
\begin{subfigure}{.13\textwidth}
    \centering
    \caption*{N2V}
    \includegraphics[width=0.98\textwidth]{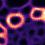}
\end{subfigure}
\begin{subfigure}{.13\textwidth}
    \centering
    \caption*{Bootstrap Hist.}
    \includegraphics[width=0.98\textwidth]{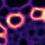}
\end{subfigure}
\begin{subfigure}{.13\textwidth}
    \centering
    \caption*{Bootstrap GMM}
    \includegraphics[width=0.98\textwidth]{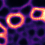}
\end{subfigure}
\begin{subfigure}{.13\textwidth}
    \centering
    \caption*{\textbf{Uncalibrated}}
    \includegraphics[width=0.98\textwidth]{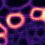}
\end{subfigure}
\begin{subfigure}{.13\textwidth}
    \centering
    \caption*{Ground Truth}
    \includegraphics[width=0.98\textwidth]{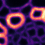}
\end{subfigure} \\
\rotatebox[origin=c]{90}{Mouse nuclei}
\begin{subfigure}{.13\textwidth}
    \centering
    \includegraphics[width=0.98\textwidth]{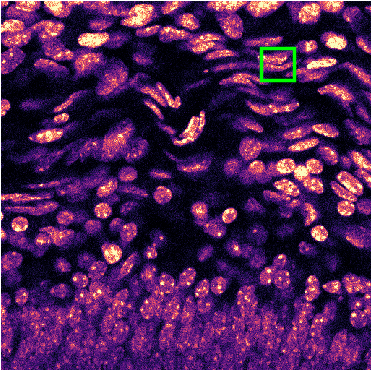}
\end{subfigure}
\begin{subfigure}{.13\textwidth}
    \centering
    \includegraphics[width=0.98\textwidth]{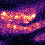}
\end{subfigure}
\begin{subfigure}{.13\textwidth}
    \centering
    \includegraphics[width=0.98\textwidth]{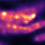}
\end{subfigure}
\begin{subfigure}{.13\textwidth}
    \centering
    \includegraphics[width=0.98\textwidth]{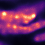}
\end{subfigure}
\begin{subfigure}{.13\textwidth}
    \centering
    \includegraphics[width=0.98\textwidth]{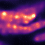}
\end{subfigure}
\begin{subfigure}{.13\textwidth}
    \centering
    \includegraphics[width=0.98\textwidth]{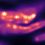}
\end{subfigure}
\begin{subfigure}{.13\textwidth}
    \centering
    \includegraphics[width=0.98\textwidth]{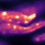}
\end{subfigure} \\
\rotatebox[origin=c]{90}{Mouse actin}
\begin{subfigure}{.13\textwidth}
    \centering
    \includegraphics[width=0.98\textwidth]{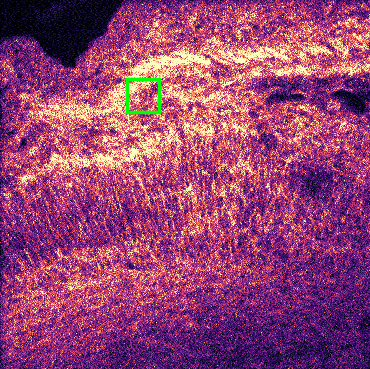}
\end{subfigure}
\begin{subfigure}{.13\textwidth}
    \centering
    \includegraphics[width=0.98\textwidth]{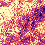}
\end{subfigure}
\begin{subfigure}{.13\textwidth}
    \centering
    \includegraphics[width=0.98\textwidth]{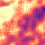}
\end{subfigure}
\begin{subfigure}{.13\textwidth}
    \centering
    \includegraphics[width=0.98\textwidth]{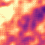}
\end{subfigure}
\begin{subfigure}{.13\textwidth}
    \centering
    \includegraphics[width=0.98\textwidth]{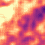}
\end{subfigure}
\begin{subfigure}{.13\textwidth}
    \centering
    \includegraphics[width=0.98\textwidth]{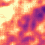}
\end{subfigure}
\begin{subfigure}{.13\textwidth}
    \centering
    \includegraphics[width=0.98\textwidth]{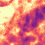}
\end{subfigure}
\caption[short]{\label{fig:images}Visual comparison of results obtained by our proposed method and the fully unsupervised methods listed in Table \ref{table:results}.  Noisy test images taken from PPN2V benchmark datasets \cite{prakash2019fully}.}
\end{figure*}

%% file: table-gaussian-params.tex
\begin{table*}[t]
\centering
\begin{tabular}{c | c c c | c c c } 
 \hline
 & Ground Truth & Estimated & G-Estimated & Ground Truth & Estimated & G-Estimated \\
 $\sigma$ & \multicolumn{3}{c|}{$a$}  & \multicolumn{3}{c}{$b$} \\ [0.5ex] 
 \hline\hline
 10 & 0 & 0.00246 & 0 & 0.00153 & 0.00148 & 0.00165 \\ 
 20 & 0 & 0.00434 & 0 & 0.00615 & 0.00605 & 0.00636 \\ 
 30 & 0 & 0.00661 & 0 & 0.0138 & 0.0136 & 0.0141 \\ 
 40 & 0 & 0.00879 & 0 & 0.0246 & 0.0244 & 0.0250 \\ 
 50 & 0 & 0.00113 & 0 & 0.0384 & 0.0381 & 0.0389 \\ 
[1ex]
 \hline
\end{tabular}
\caption[short]{\label{table:gaussian-params}Estimated $a$,$b$ parameters for varying levels of Gaussian noise using our Poisson-Gaussian and pure Gaussian noise fitting techniques.}
\end{table*}

%% file: table-gaussian-psnr.tex
\begin{table*}[t]
\centering
\begin{tabular}{ c | c | c | c | c} 
 \hline
 & Pseudo-Clean & Uncalibrated & Uncalibrated w/ $a=0$ & Ground Truth \\
 $\sigma$ & PSNR & PSNR  & PSNR & PSNR \\ [0.5ex] 
 \hline\hline
 10 & 39.25 & 39.50 & 39.58 & 39.61 \\ 
 20 & 36.73 & 36.84 & 36.96 & 37.02 \\ 
 30 & 35.04 & 35.10 & 35.26 & 35.30 \\ 
 40 & 33.72 & 33.76 & 33.92 & 33.95 \\ 
 50 & 32.74 & 32.79 & 32.91 & 32.94 \\ 
[1ex]
 \hline
\end{tabular}
\caption[short]{\label{table:gaussian-psnr}PSNR comparison between using our uncalibrated method with Poisson-Gaussian noise fitting and Gaussian noise fitting for estimating the $a$,$b$ parameters of pure Gaussian noise.}
\end{table*} 

%% file: table-poisson-params.tex
\begin{table*}[t]
\centering
\begin{tabular}{c | c c c | c c c } 
 \hline
 & Ground Truth & Estimated & P-Estimated & Ground Truth & Estimated & P-Estimated \\
 $\lambda$ & \multicolumn{3}{c|}{$a$}  & \multicolumn{3}{c}{$b$} \\ [0.5ex] 
 \hline\hline
 10 & 0.100 & 0.108 & 0.100 & 0 & -0.000390 & 0 \\ 
 20 & 0.0500 & 0.0561 & 0.0500 & 0 & -0.000246 & 0 \\ 
 30 & 0.0333 & 0.0380 & 0.0348 & 0 & -0.000148 & 0 \\ 
 40 & 0.0250 & 0.0290 & 0.0258 & 0 & -0.000152 & 0 \\ 
 50 & 0.0200 & 0.0239 & 0.0208 & 0 & -0.000142 & 0 \\ 
[1ex]
 \hline
\end{tabular}
\caption[short]{\label{table:poisson-params}Estimated $a$,$b$ parameters for varying levels of Poisson noise using our Poisson-Gaussian and pure Poisson noise fitting techniques.}
\end{table*}

%% file: table-poisson-psnr.tex
\begin{table*}[t]
\centering
\begin{tabular}{ c | c | c | c | c} 
 \hline
 & Pseudo-Clean & Uncalibrated & Uncalibrated w/ $b=0$ & Ground Truth \\
 $\lambda$ & PSNR & PSNR  & PSNR & PSNR \\ [0.5ex] 
 \hline\hline
 10 & 34.15 & 34.29 & 34.38 & 34.38 \\ 
 20 & 35.66 & 35.76 & 35.90 & 35.93 \\ 
 30 & 36.61 & 36.71 & 36.82 & 36.89 \\ 
 40 & 37.09 & 37.22 & 37.36 & 37.40 \\ 
 50 & 37.58 & 37.71 & 37.83 & 37.91 \\ 
[1ex]
 \hline
\end{tabular}
\caption[short]{\label{table:poisson-psnr}PSNR comparison between using our uncalibrated method with Poisson-Gaussian noise fitting and Poisson noise fitting for estimating the $a$,$b$ parameters of pure Poisson noise.}
\end{table*} 

%% file: table-synthetic-params-full.tex
\begin{table*}[t]
\centering
\begin{tabular}{c | c | c c c | c c c} 
 \hline
 & & Ground Truth & Estimated & Absolute Error & Ground Truth & Estimated & Absolute Error\\
 $\lambda$ & $\sigma$ & \multicolumn{3}{c|}{$a$}  & \multicolumn{3}{c}{$b$} \\ [0.5ex] 
 \hline\hline
  & 10 & 0.100 & 0.105 & 0.00537 & 0.00153 & 0.00116 & 0.000376 \\ 
  & 20 & 0.100 & 0.103 & 0.00315 & 0.00615 & 0.00645 & 0.000296 \\ 
 10 & 30 & 0.100 & 0.114 & 0.0135 & 0.0138 & 0.0133 & 0.000539 \\ 
  & 40 & 0.100 & 0.121 & 0.0209 & 0.0246 & 0.0234 & 0.00122 \\ 
  & 50 & 0.100 & 0.117 & 0.0177 & 0.0384 & 0.0379 & 0.000578 \\ 
 \hline\hline
  & 10 & 0.0500 & 0.0565 & 0.00651 & 0.00153 & 0.00128 & 0.000254 \\ 
  & 20 & 0.0500 & 0.0593 & 0.00925 & 0.00615 & 0.00579 & 0.000356 \\ 
 20 & 30 & 0.0500 & 0.0608 & 0.0108 & 0.0138 & 0.0134 & 0.000388 \\ 
  & 40 & 0.0500 & 0.0623 & 0.0123 & 0.0246 & 0.0241 & 0.000405 \\ 
  & 50 & 0.0500 & 0.0658 & 0.0158 & 0.0384 & 0.0378 & 0.000614 \\ 
 \hline\hline
  & 10 & 0.0333 & 0.0392 & 0.00587 & 0.00153 & 0.00134 & 0.000193 \\ 
  & 20 & 0.0333 & 0.0408 & 0.00751 & 0.00615 & 0.00590 & 0.000247 \\ 
 30 & 30 & 0.0333 & 0.0430 & 0.00975 & 0.0138 & 0.0134 & 0.000387 \\ 
  & 40 & 0.0333 & 0.0457 & 0.0123 & 0.0246 & 0.0241 & 0.000504 \\ 
  & 50 & 0.0333 & 0.0477 & 0.0144 & 0.0384 & 0.0378 & 0.000623 \\ 
 \hline\hline
  & 10 & 0.0250 & 0.0305 & 0.00551 & 0.00153 & 0.00131 & 0.000228 \\ 
  & 20 & 0.0250 & 0.0325 & 0.00751 & 0.00615 & 0.00586 & 0.000291 \\ 
 40 & 30 & 0.0250 & 0.0340 & 0.00897 & 0.0138 & 0.0135 & 0.000312 \\ 
  & 40 & 0.0250 & 0.0367 & 0.0117 & 0.0246 & 0.0242 & 0.000405 \\ 
  & 50 & 0.0250 & 0.0401 & 0.0151 & 0.0384 & 0.0377 & 0.000352 \\ 
 \hline\hline
  & 10 & 0.0200 & 0.0250 & 0.00508 & 0.00153 & 0.00135 & 0.000184 \\ 
  & 20 & 0.0200 & 0.0267 & 0.00675 & 0.00615 & 0.00591 & 0.000288 \\ 
 50 & 30 & 0.0200 & 0.0281 & 0.00807 & 0.0138 & 0.0137 & 0.000180 \\ 
  & 40 & 0.0200 & 0.0297 & 0.00969 & 0.0246 & 0.0244 & 0.000198 \\ 
  & 50 & 0.0200 & 0.0329 & 0.0129 & 0.0384 & 0.0381 & 0.000353 \\ 
[1ex]
 \hline
\end{tabular}
\caption[short]{\label{table:synthetic-params-full}Quantitative comparison of fitting a Poisson-Gaussian noise model on all the different Poisson-Gaussian noise levels of our synthetic Confocal MICE dataset.}
\end{table*}

%% file: table-synthetic-psnr-full.tex
\begin{table*}[t]
\centering
\begin{tabular}{c | c | c | c | c } 
 \hline
 & & Pseudo-Clean & Uncalibrated & Ground Truth \\
 $\lambda$ & $\sigma$ & PSNR & PSNR  & PSNR \\ [0.5ex] 
 \hline\hline
  & 10 & 33.96 & 34.17 & 34.20 \\ 
  & 20 & 33.60 & 33.75 & 33.82 \\ 
 10 & 30 & 33.09 & 33.14 & 33.27 \\ 
  & 40 & 32.38 & 32.42 & 32.55 \\ 
  & 50 & 31.98 & 32.03 & 32.14 \\ 
 \hline\hline
  & 10 & 35.45 & 35.56 & 35.73 \\ 
  & 20 & 34.68 & 34.75 & 34.93 \\ 
 20 & 30 & 33.97 & 34.03 & 34.19 \\ 
  & 40 & 33.17 & 33.22 & 33.37 \\ 
  & 50 & 32.22 & 32.27 & 32.39 \\ 
 \hline\hline
  & 10 & 36.18 & 36.26 & 36.43 \\ 
  & 20 & 35.30 & 35.37 & 35.56 \\ 
 30 & 30 & 34.10 & 34.15 & 34.33 \\ 
  & 40 & 33.16 & 33.22 & 33.37 \\ 
  & 50 & 32.26 & 32.34 & 32.45 \\ 
 \hline\hline
  & 10 & 36.61 & 36.71 & 36.91 \\ 
  & 20 & 35.45 & 35.53 & 34.71 \\ 
 40 & 30 & 34.47 & 34.52 & 34.33 \\ 
  & 40 & 33.30 & 33.37 & 33.52 \\ 
  & 50 & 31.73 & 31.78 & 31.92 \\ 
 \hline\hline
  & 10 & 36.95 & 37.05 & 37.25 \\ 
  & 20 & 35.83 & 35.90 & 36.10 \\ 
 50 & 30 & 34.47 & 34.54 & 34.73 \\ 
  & 40 & 33.52 & 33.57 & 33.72 \\ 
  & 50 & 32.54 & 32.60 & 32.74 \\ 
 [1ex]
 \hline
\end{tabular}
\caption[short]{\label{table:synthetic-psnr-full}Comparison of PSNR using our bootstrapping method on all the different Poisson-Gaussian noise levels of our synthetic Confocal MICE dataset.}
\end{table*} 

%% file: table-psnr-param-error.tex
\begin{table*}[t]
\centering
\begin{tabular}{c | c | c | c | c } 
 \hline
 $\%$ error & $\%$ error & Pseudo-Clean & Uncalibrated & Ground Truth \\
 $a$ & $b$ & PSNR & PSNR  & PSNR \\ [0.5ex] 
 \hline\hline
  & 0 & 34.10 & 34.28 & 34.33 \\ 
 6 & 4 & 34.10 & 34.31 & 34.33 \\ 
  & 8 & 34.10 & 34.30 & 34.33 \\ 
 \hline\hline
  & 0 & 34.10 & 34.17 & 34.33 \\ 
 22 & 4 & 34.10 & 34.19 & 34.33 \\ 
  & 8 & 34.10 & 34.24 & 34.33 \\ 
 \hline\hline
  & 0 & 34.10 & 34.13 & 34.33 \\ 
 38 & 4 & 34.10 & 34.14 & 34.33 \\ 
  & 8 & 34.10 & 34.15 & 34.33 \\ 
 [1ex]
 \hline
\end{tabular}
\caption[short]{\label{table:psnr-param-error}Quantitative results on how our uncalibrated method's PSNR is affected by varying percent errors in our estimated $a$,$b$ parameters. The results shown were done using Poisson-Gaussian noise with $\lambda=30$ and $\sigma=30$. \\ [1ex]}
\end{table*} 

%% file: table-bsd68.tex
\begin{table*}[t]
\centering
\begin{tabular}{c | c | c | c | c | c | c | c} 
 \hline
 & & & & & & & Uncalibrated (Ours) \\
 & BM3D & Traditional & N2N & N2V & Pseudo-Clean (Ours) & Uncalibrated (Ours) & w/ known b \\ [0.5ex] 
 \hline\hline
 BSD68 & 28.59 & 29.06 & 28.86 & 27.71 & 26.95 & 27.80 & 28.15 \\ 
[0.5ex]
 \hline
\end{tabular}
\caption[short]{\label{table:bsd68}PSNR results on the BSD68 dataset.}
\end{table*} 

%% file: table-bsd68-params.tex
\begin{table*}[t]
\centering
\begin{tabular}{c | c c | c c } 
 \hline
 & Ground Truth & Estimated & Ground Truth & Estimated \\
 $\sigma$ & \multicolumn{2}{c|}{$a$}  & \multicolumn{2}{c}{$b$} \\ [0.5ex] 
 \hline\hline
 25 & 0 & 0.00533 & 0.00961 & 0.00861 \\ 
[1ex]
 \hline
\end{tabular}
\caption[short]{\label{table:bsd68-params}Estimated noise parameters using our uncalibrated method on the BSD68 dataset.}
\end{table*} 